\documentclass{aa}

\usepackage{graphicx}
\usepackage{epstopdf}
\usepackage{txfonts}
\usepackage{graphicx,url,twoopt,natbib}
\usepackage[breaklinks=true,backref=page]{hyperref} %% to avoid \citeads line fills \bibpunct{(}{)}{;}{a}{}{,} %% natbib format for A&A and ApJ \makeatletter \newcommandtwoopt{\citeads}[3][][]{\href{http://adsabs.harvard.edu/abs/#3}% {\def\hyper@linkstart##1##2{}% \let\hyper@linkend\@empty\citealp[#1][#2]{#3}}} \newcommandtwoopt{\citepads}[3][][]{\href{http://adsabs.harvard.edu/abs/#3}% {\def\hyper@linkstart##1##2{}% \let\hyper@linkend\@empty\citep[#1][#2]{#3}}} \newcommandtwoopt{\citetads}[3][][]{\href{http://adsabs.harvard.edu/abs/#3}% {\def\hyper@linkstart##1##2{}% \let\hyper@linkend\@empty\citet[#1][#2]{#3}}} \newcommandtwoopt{\citeyearads}[3][][]% {\href{http://adsabs.harvard.edu/abs/#3} {\def\hyper@linkstart##1##2{}% \let\hyper@linkend\@empty\citeyear[#1][#2]{#3}}}

\hypersetup{
  colorlinks=true,   %% links colored instead of frames
  urlcolor=green,     %% external hyperlinks
  linkcolor=red,     %% internal latex links (eg Fig)
  citecolor=blue,
}

\makeatother

\usepackage{amstext}
\usepackage[normalem]{ulem}

\begin{document}

   \title{Subdwarf B stars from the common envelope ejection channel}

%   \subtitle{I. Overviewing the $\kappa$-mechanism}

   \author{H. Xiong,
          \inst{1,2,3,4}
          \,
          X. Chen,
          \inst{1,2,3}
          \,
          Ph. Podsiadlowski,
          \inst{5}
           Y. Li
          \inst{1,2,3}
          \and
           Z. Han
          \inst{1,2,3}
%          \fnmsep\thanks{}
          }

   \institute{Yunnan Observatories, Chinese Academy of Sciences, 396 Yangfangwang, Guandu District, Kunming, 650216, P. R. China\\
              \email{cxf@ynao.ac.cn}\\
              \email{zhanwenhan@ynao.ac.cn}\\
         \and
               Key Laboratory for the Structure and Evolution of Celestial Objects, Chinese Academy of Sciences, 396 Yangfangwang, Guandu District, Kunming, 650216, P. R. China\\
         \and
              Center for Astronomical Mega-Science, Chinese Academy of Sciences, 20A Datun Road, Chaoyang District, Beijing, 100012, P. R. China\\
         \and
              University of Chinese Academy of Sciences, Beijing 100049, China\\
         \and
              Department of Astrophysics, University of Oxford, Oxford OX1 3RH, UK\\
             }

%   \date{Received ...; accepted ...}

% \abstract{}{}{}{}{}
% 5 {} token are mandatory

  \abstract
{Subdwarf B stars (sdB) are important to stellar evolutionary
theory and asteroseismology, and they are crucial to our understanding
of the structure and evolution of the Galaxy. According to the canonical
binary scenario, the majority of sdBs are produced from low-mass
stars with degenerate cores where helium is ignited in a way of
flashes. Owing to numerical difficulties, the models of produced sdBs are
generally constructed from more massive stars with non-degenerate
cores. This leaves several uncertainties on the exact
characteristics of sdB stars. }
{The purpose of this paper is to study the characteristics of sdBs
produced from the common envelope (CE) ejection channel.}
{We used the stellar evolution code {\it \textup{Modules for Experiments in
Stellar Astrophysics}} (MESA), which can resolve flashes during
stellar evolution. To mimic the CE ejection process, we first
evolved a single star to a position near the tip of the red giant
branch, then artificially removed its envelope with a very high
mass-loss rate until the envelope began to shrink. Finally, we
followed the evolution of the remnant until it became a helium
or a carbon-oxygen white dwarf.}
{The sdB stars produced from the CE ejection channel appear to form
two distinct groups on the effective temperature-gravity diagram.
One group, referred to as the flash-mixing sdBs, almost has no H-rich
envelope and crowds at the hottest temperature end of the extreme
horizontal branch (EHB), while the other group, called the
canonical sdBs, has significant H-rich envelope and is spread throughout
the entire canonical EHB region. The key factor for the dichotomy
of the sdB properties is the development of convection during the
first helium flash, that is, that the convection region penetrates the
H-rich envelope in the case of the flash-mixing sdBs, but fails
to do this in the case of the
canonical sdBs. }
{The dichotomy of the sdB properties from the CE ejection channel
is intrinsic and caused by the interior structure of the star
after the CE ejection. The modelling of the CE ejection process
will greatly change the parameter space for the two typical groups of sdB stars.
If the CE ejection stops early for a given initial stellar mass and a given core mass at the onset of the CE,
then the star has a relatively massive H-rich
envelope, which generally results in a canonical sdB. Observationally, only a few sdB binaries with short
orbital periods are located in the flash-mixing sdB region, and
there is a lack of He-rich sdBs
in binaries with short orbital periods. This indicates that flash mixing is
not very frequent in products of the CE ejection. A falling-back
process after the CE ejection, similar to what occurs in nova,
is an appropriate way of increasing the envelope mass, and it
then
prevents flash mixing.}


   \maketitle
%
%________________________________________________________________

\section{Introduction}

In the Hertzsprung-Russell diagram (HRD), subdwarf B (sdB) stars
are located between the upper main sequence (MS) and the white
dwarf (WD) cooling sequence at the blueward extension of the
horizontal branch. They are also known as extreme
horizontal-branch (EHB) stars in globular clusters (GCs). SdB
stars are important in several aspects. The study of their origin
significantly improved our knowledge of stellar and binary
evolution theory  \citep[see the review of][]{Heber2009,Heber2016}. Short-period sdB binary systems are
candidates of type Ia supernova progenitors
\citep{Maxted2000a,Wang}. Many sdB stars show
multiperiodic pulsations, and those stars are important objects of
asteroseismology study \citep {Charpinet}. SdB stars have
been used as distance indicators and as a probe to study the
Galactic structure and evolution \citep[see the review of][]{Heber2009,Heber2016,Altmann}. They are also considered to be crucial
sources of far-ultraviolet radiation in early-type galaxies
 \citep {Ferguson,Brown2000,Han2007} since they are hot (with an effective temperature
$T_{\rm eff}$ of between 20000-40000 K) and have relatively long
lifetimes ($\sim 10^{8} {\rm yr}$).

SdB stars are generally believed to be helium-core-burning stars
with extremely thin hydrogen envelopes( $<0.02 M_\odot$). More
than half of them are found in binaries { \citep {Maxted2001,Napiwotzki,Copperwheat}. \cite{Han2002,Han2003} developed a detailed binary model for the
formation of sdBs that successfully explains field sdBs,
possibly EHB stars in GCs \citep {Han2008}, and sdBs on long-orbital periods observed recently \citep{Chen2013}. In the binary scenario,
there are three formation channels for sdBs: stable Roche-lobe overflow (RLOF) for those with long orbital periods, common
envelope (CE) ejection for those with short orbital periods, and
the merger of helium white dwarfs (WD) for single sdBs.

The key point of the three channels is the He ignition in
the core. In most instances, the He core is degenerate and the He
ignition is explosive, that is, there are several He flashes before stable He core burning is established. Because of numerical
difficulties in dealing with flashes in most stellar evolution
codes, almost all the sdB models are artificially constructed from
relatively massive stars with non-degenerate cores   \citep[e.g.][] {Brown2001,Han2002}. The constructed sdB stars generally
have the same core mass as in the progenitors at the He
ignition, but the composition in the envelope may be different based on
different assumptions. The He flashes may alter envelope mass and
element abundances  \citep {Brown2001,Sweigart}. A
detailed study of the He flash process before the stable He-core
burning is necessary and important for our understanding of the
characteristics of sdB stars and for the study of
asteroseismology. The newly developed stellar evolution code MESA
\citep[Modules for Experiments in Stellar Astrophysics,][]{Paxton2011,Paxton2013,Paxton2015} can resolve dramatic changes such as
flashes in stellar evolution and therefore provides a good opportunity for
studying the characteristics of sdBs in details. In general,
the basic equations of stellar structure and evolution are written as difference equations
in stellar evolution codes. The dramatic changes in stellar structure such as flashes make the difference approximation invalid and the codes are then difficult to converge.
MESA employs adaptive mesh refinement and sophisticated timestep controls,
and the convergence criteria will be automatically adjusted under some difficult circumstances such as flashes.
All these factors facilitate using MESA  to deal with the dramatic changes in stellar structures without  the convergence problem
\citep{Paxton2011,Paxton2013}.

\cite{Schindler} constructed a series of sdB models using
MESA. These models reproduced the general properties of the
zero-age EHB and the interior structures of sdB stars from
asteroseismology. The origin of sdB stars is ignored in their
study, which may play a crucial role in the structure of sdB
stars, especially for the mass and composition in the envelope.
For example, sdB stars from stable RLOF are considered to have
a higher envelope mass than those from the CE ejection channel
\citep{Han2002}. This envelope mass may further affect the He flash
process and the final envelope mass and composition of sdB stars.
Much evidence shows that most of sdBs are in short orbit-period
binaries \citep{Saffer1998,Jeffery,Koen,Orosz,Moran,Maxted2000a,Maxted2000b,Maxted2001,Barlow,Kupfer}. This means that the CE ejection channel is a
major mechanism for producing sdBs.

In this paper, we employ MESA to
systematically study the characteristics of sdBs from the
CE ejection channel. In Sect. 2 we introduce the basic
inputs in the code and the method of simulating the CE process.
The results are presented in Sect. 3, where we describe that
the sdB stars produced from the CE ejection channel show two distinct groups
on the effective temperature-gravity diagram. In Sect. 4
we discuss how the results are influenced by the treatment of CE ejection,
mass loss during He flashes, metallicity, and some observationally related objects.
A conclusion is given in Sect. 5.
\begin{figure}
\includegraphics[width=8cm]{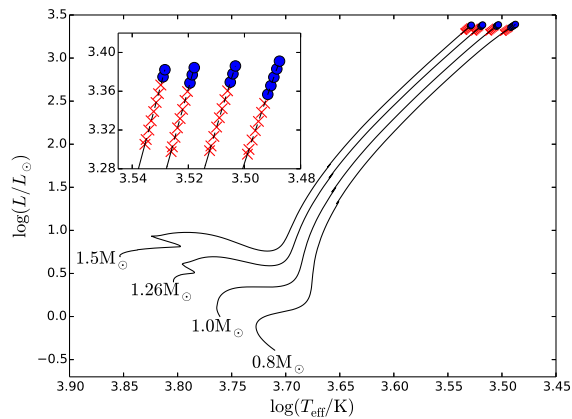}
\caption{Evolutionary tracks on the Hertzsprung-Russell diagram
for the four stars studied in the paper. The X and the filled
dot mark the positions where the star fills its Roche lobe
and begins the CE evolution. The X shows stars with a strong H flash after the main He flash and
the filled dots stars without such a strong H flash (see text for
details).
\label{fig 1}}%
\end{figure}

\begin{figure*}
\centering
   %%%
\includegraphics[width=15.9cm]{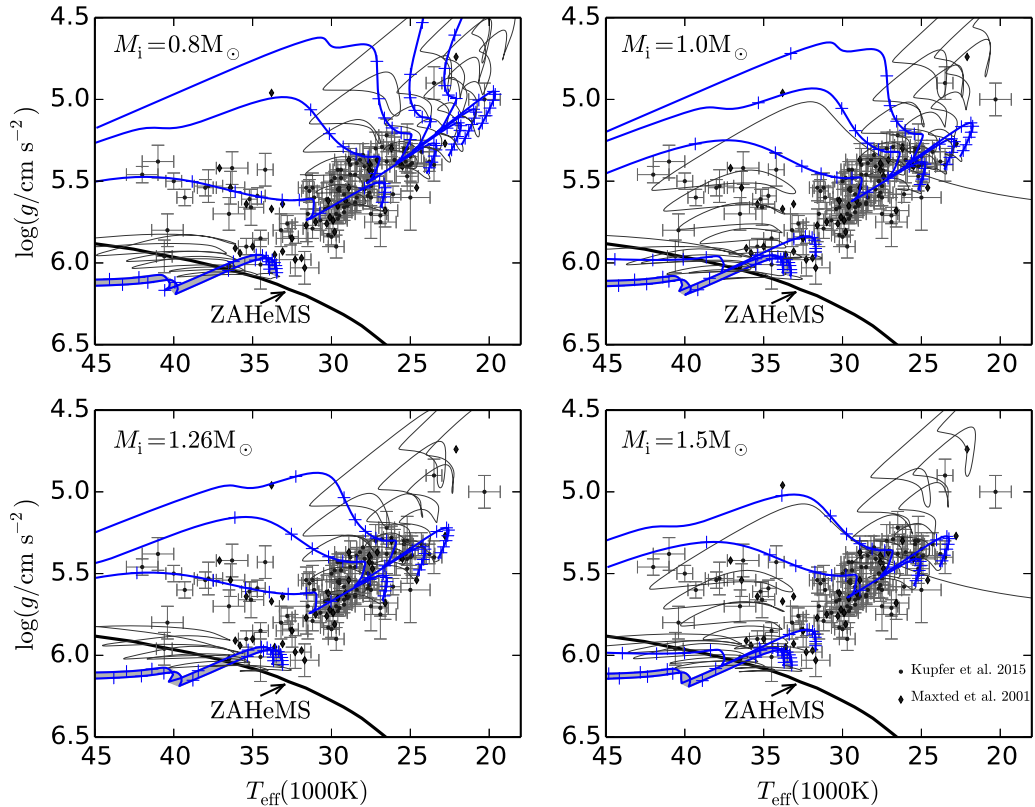}
\caption{Evolutionary tracks of the produced sdB stars on the
effective temperature-gravity ($T_{\rm eff}-{\rm log} g$) diagram.
Initial stellar mass, $M_{\rm i}$, and the position of zero-age helium main
sequence (ZAHeMS, the dot-dashed line) are indicated.
The shaded area between the lowest two lines in each panel shows
the region of the evolutionary tracks of the flash-mixing models
(see the text for details) for clarity. The
age difference between adjacent crosses is $10^7 {\rm yr}$.
The dots and diamonds are sdBs with short
orbital periods from \cite{Maxted2001,Kupfer}, respectively. The grey
lines are the tracks during helium flashes before stable He-core
burning.
\label{fig 2}}%
\end{figure*}
\section{Stellar evolution calculations}

We used version 7184 of MESA and adopted the physics options
similar to those in the standard model of \cite {Schindler},
that is, the element abundances are chosen for population I stars, $Z=0.02$ and $X=0.70$ for the metallicity and hydrogen
mass fraction, respectively, and the nuclear network is
$pp\underline{\makebox[1em]{}}cno\underline{\makebox[1em]{}}
extras\underline{\makebox[1em]{}}o18\underline{\makebox[1em]{}}ne22.net$,
in which all relevant reactions for H and He burning are included. The
mixing length parameter, $\alpha_{\rm{MLT}}$, is set to 2 and the
opacity table is OPAL type II, which allows the abundances of C
and O to vary with time. For simplicity, no stellar wind or other
mass loss is included in our calculations except for the CE
ejection.

In the CE ejection channel, the progenitor of an sdB is a giant
and fills its Roche lobe near the tip of red giant branch (RGB).
The following mass transfer is dynamically unstable, and a common
envelope forms. The donor core and the companion spiral in the
CE. As a result of the friction between the inner binary and the CE, the
orbit decays and the orbital energy is released and deposited in
the envelope. The envelope may be ejected eventually when the
released orbital energy is higher than the binding energy of the
envelope  \citep{Han2002}. The whole process is
dynamical
\citep[with a timescale of $\sim 10^3$ yr, see][ and references therein]{Ivanova}, and cannot yet be simulated by MESA.
We therefore modelled this process in a way similar to that of
\cite{Han2002}. We first evolved a single star to the
position where the CE begins (near the tip of RGB),
then artificially removed the envelope with a high mass-loss rate ($10^{-3}M_\odot{\rm
yr^{-1}}$). In general, the giant expands dramatically
as a result of mass loss and suddenly contracts after almost all of the envelope has been lost.
We stopped the mass loss when the star started to collapse,
that is, when the radius of the star, $R$, was equal to the initial radius at the beginning of mass loss, $R_{\rm 0}$.
The evolution of the remnant was followed until the surface temperature is lower than 5000 K,
that is, the star evolved to a cool He WD or a carbon-oxygen WD (after an sdB phase).
 The treatment of the CE process may affect the properties of the produced sdBs,
as discussed in Sect. 4.

We adopted four initial stellar masses (the mass donor) for our study: $M_{\rm i}=0.8M_\odot$, 1.0$M_\odot$, 1.26$M_\odot,$ and
1.5$M_\odot$. For each mass, we systematically
investigated a series of positions where the CE begins on the RGB,
that is, the core mass increases from the minimum mass
allowed for He ignition
to the tip of RGB in steps of 0.002$M_\odot$. If the core mass at the onset of the CE
is lower than the minimum mass, helium cannot be ignited and the remnant will directly evolve into a He WD  \citep[see][]{Han2002}. Figure\,\ref{fig 1}
shows the positions to be studied on the evolutionary tracks. The
initial mass and core mass for each point are listed in Table.\,\ref{table 1}.

\begin{table*}
\caption{Properties of sdB stars produced from the CE ejection
channel in our study. The first four columns show the initial
mass of the donors $M_{\rm i}$, the core mass at the onset of the
CE process $M_{\rm c}$, the core mass just before the first He flash $M_{\rm c}^*$, and the remnant mass after the CE process
$M_{\rm sdB}$, respectively. Columns 5-9 list the total
hydrogen mass $M_{\rm H}$, surface He abundance ($Y$), surface
C abundance ($X_{\rm C}$), surface N abundance ($X_{\rm N}$), and surface
O abundance ($X_{\rm O}$), respectively, as He burns stably in the core. The masses are in units of solar mass.
 Note that the differences between $M_{\rm c}$ and $M_{\rm c}^*$ are caused by
the H-shell burning after the CE ejection. The treatment of the CE ejection process may affect the envelope mass after the ejection and the core mass at the first He flash. However, the main results will not change, as discussed in Sect. 4.} \label{table 1}
\centering
\begin{tabular}{cccccccccc}     % 8 columns
\hline\hline
 $M_{\rm i}$ & $M_{\rm c}$& $M_{\rm c}^*$  &$M_{\rm sdB}$  &$M_{\rm H}$&Y&$X_{\rm C}$&$X_{\rm N}$&$X_{\rm O}$\\
 \hline
  & 0.443  &0.452  &0.453  &$1.068\times10^{-4}$&0.947 &0.020&0.015&$7.891\times10^{-4}$\\
  & 0.444  &0.453  &0.454  &$1.593\times10^{-4}$&0.943 &0.021&0.014&$8.033\times10^{-4}$\\
  & 0.446  &0.455  &0.456  &$0.973\times10^{-4}$&0.948 &0.023&0.012&$6.717\times10^{-4}$\\
  & 0.448  &0.458  &0.458  &$1.714\times10^{-4}$&0.941 &0.025&0.010&$7.323\times10^{-4}$\\
  & 0.45  &0.460   &0.461  &$3.724\times10^{-4}$&0.925 &0.027&0.008&$9.798\times10^{-4}$\\
 0.8 & 0.452  &0.462  &0.463  &$1.100\times10^{-4}$&0.945 &0.029&0.007&$7.146\times10^{-4}$\\
  & 0.454   &0.464  &0.465  &$0.991\times10^{-4}$&0.943 &0.034&0.003&$9.053\times10^{-4}$\\
  & 0.456   &0.466  &0.467  &$0.733\times10^{-3}$&0.298 &0.003&0.001&$9.357\times10^{-3}$\\
  & 0.458   &0.466  &0.470  &$2.589\times10^{-3}$&0.298 &0.003&0.001&$9.357\times10^{-3}$\\
  & 0.46   &0.466   &0.472  &$4.054\times10^{-3}$&0.298 &0.003&0.001&$9.357\times10^{-3}$\\
  & 0.462   &0.466  &0.473  &$5.129\times10^{-3}$&0.298 &0.003&0.001&$9.357\times10^{-3}$\\
  & 0.464   &0.466  &0.476  &$6.855\times10^{-3}$&0.298 &0.003&0.001&$9.357\times10^{-3}$\\
  &&&\\
% \hline
  & 0.443   &0.449  &0.450  &$1.359\times10^{-4}$&0.944 &0.019&0.016&$1.133\times10^{-3}$\\
  & 0.444   &0.450  &0.451  &$2.994\times10^{-4}$&0.931 &0.021&0.014&$1.061\times10^{-3}$\\
  & 0.446   &0.453  &0.453  &$1.113\times10^{-4}$&0.942 &0.022&0.013&$7.911\times10^{-4}$\\
  & 0.448   &0.455  &0.455  &$1.383\times10^{-4}$&0.939 &0.023&0.011&$7.524\times10^{-4}$\\
  & 0.45   &0.457   &0.457  &$0.984\times10^{-4}$&0.943 &0.026&0.009&$7.830\times10^{-4}$\\
 1.0 & 0.452   &0.459  &0.459  &$2.387\times10^{-6}$&0.943 &0.028&0.008&$7.086\times10^{-4}$\\
  & 0.454   &0.461  &0.462  &$1.316\times10^{-4}$&0.942 &0.031&0.005&$6.969\times10^{-4}$\\
  & 0.456   &0.463  &0.463  &$3.110\times10^{-4}$&0.759 &0.018&0.003&$1.049\times10^{-2}$\\
  & 0.458   &0.463  &0.466  &$1.099\times10^{-3}$&0.301 &0.003&0.001&$9.356\times10^{-3}$\\
  & 0.46   &0.464   &0.468  &$2.061\times10^{-3}$&0.301 &0.003&0.001&$9.356\times10^{-3}$\\
  & 0.462   &0.464  &0.470  &$2.849\times10^{-3}$&0.301 &0.003&0.001&$9.356\times10^{-3}$\\
 &&&\\
  & 0.443   &0.448  &0.449  &$1.992\times10^{-4}$&0.938 &0.019&0.016&$1.089\times10^{-3}$\\
  & 0.444   &0.449  &0.450 &$1.663\times10^{-4}$&0.942 &0.020&0.014&$8.811\times10^{-4}$\\
  & 0.446   &0.451  &0.452  &$2.021\times10^{-4}$&0.938 &0.023&0.012&$9.771\times10^{-4}$\\
  & 0.448   &0.453  &0.454  &$1.982\times10^{-4}$&0.941 &0.024&0.011&$8.197\times10^{-4}$\\
  & 0.45   &0.455   &0.456  &$3.228\times10^{-4}$&0.928 &0.025&0.010&$9.324\times10^{-4}$\\
 1.26 & 0.452   &0.457 &0.458  &$1.782\times10^{-4}$&0.940 &0.027&0.009&$7.437\times10^{-4}$\\
  & 0.454   &0.459  &0.459  &$1.539\times10^{-4}$&0.941 &0.029&0.007&$7.031\times10^{-4}$\\
  & 0.456   &0.461  &0.461  &$1.752\times10^{-4}$&0.938 &0.033&0.004&$1.195\times10^{-3}$\\
  & 0.458   &0.462  &0.463  &$0.732\times10^{-3}$&0.299 &0.003&0.002&$9.353\times10^{-3}$\\
  & 0.46   &0.462   &0.465  &$1.981\times10^{-3}$&0.299 &0.003&0.002&$9.353\times10^{-3}$\\
  & 0.462   &0.463  &0.467  &$3.192\times10^{-3}$&0.299 &0.003&0.002&$9.353\times10^{-3}$\\
 &&&\\
  & 0.445   &0.449  &0.449  &$1.406\times10^{-4}$&0.943 &0.021&0.014&$8.786\times10^{-4}$\\
  & 0.446   &0.450  &0.450  &$1.918\times10^{-4}$&0.939 &0.022&0.013&$8.543\times10^{-4}$\\
  & 0.448   &0.452  &0.452  &$3.265\times10^{-4}$&0.928 &0.023&0.012&$9.879\times10^{-4}$\\
  & 0.45   &0.454   &0.454  &$1.877\times10^{-4}$&0.940 &0.024&0.011&$7.597\times10^{-4}$\\
 1.5 & 0.452 &0.456   &0.456  &$1.949\times10^{-4}$&0.939 &0.026&0.009&$7.373\times10^{-4}$\\
  & 0.454   &0.457  &0.458  &$1.527\times10^{-4}$&0.941 &0.028&0.008&$7.539\times10^{-4}$\\
  & 0.456  &0.460   &0.460  &$1.206\times10^{-4}$&0.943 &0.031&0.006&$7.480\times10^{-4}$\\
  & 0.458  &0.462   &0.462  &$3.002\times10^{-4}$&0.758 &0.019&0.003&$1.091\times10^{-2}$\\
  & 0.46   &0.462   &0.464  &$1.343\times10^{-3}$&0.294 &0.003&0.002&$9.350\times10^{-3}$\\
  & 0.462   &0.462  &0.466  &$2.540\times10^{-3}$&0.294 &0.003&0.002&$9.350\times10^{-3}$\\
 \hline\hline
\end{tabular}
\label{table 1}
\end{table*}

\section{Characteristics of sdB Stars from our models}

\subsection{Location on the $T_{\rm eff}-{\rm log} g$ diagram}

\begin{figure}
\centering
   %%%
\includegraphics[width=8cm]{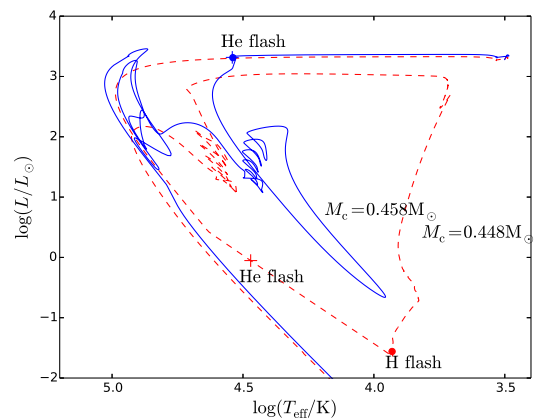}
\caption{Typical evolutionary tracks (starting from the end of the CE ejection)
for the two groups of sdBs produced from the CE ejection channel.
The two models have the
same initial masses ($M_{\rm i}=0.8M_\odot$), but different core
masses at the onset of the CE ejection, that is, $M_{\rm
c}=0.448M_\odot$ for that of the flash-mixing sdB (the dashed
line) and $0.458M_\odot$ for that of the canonical sdB (the solid
line), respectively. The plus and the filled dot show the
positions where the first He and H flash occur, respectively.
\label{fig 3}}%
\end{figure}

\begin{figure*}
   \centering
   %%%
\includegraphics[width=15cm]{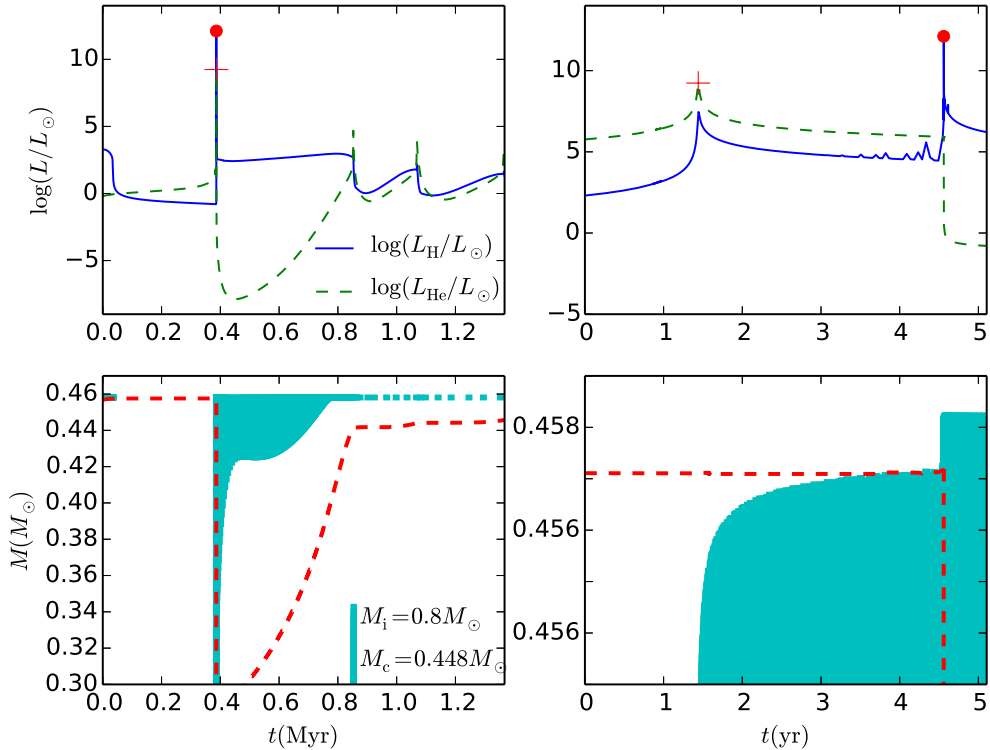}
\caption{Temporal evolution of the H- and He-burning luminosity
(the upper panels) and the structure of convection region (the
shaped region) near the surface (the bottom panels) during the
flashes. The model has $M_{\rm i}=0.8M_\odot$ and $M_{\rm
c}=0.448M_\odot$ at the onset of the CE ejection.
The plus and the filled dot show the positions
 where the first He and H flash occur, respectively (see Fig.\,\ref{fig 3}).
The dashed line in the bottom panels is the boundary of the H-rich envelope, that
is, the H mass abundance of $X_{\rm H}=0.1$.
The convection developed during the first He flash penetrates the H-rich envelope,
resulting in an additional very strong H flash.
The right two panels zoom in on the first He and H flashes.
\label{fig 4}}%
\end{figure*}

\begin{figure*}
\centering
   %%%
\includegraphics[width=15cm]{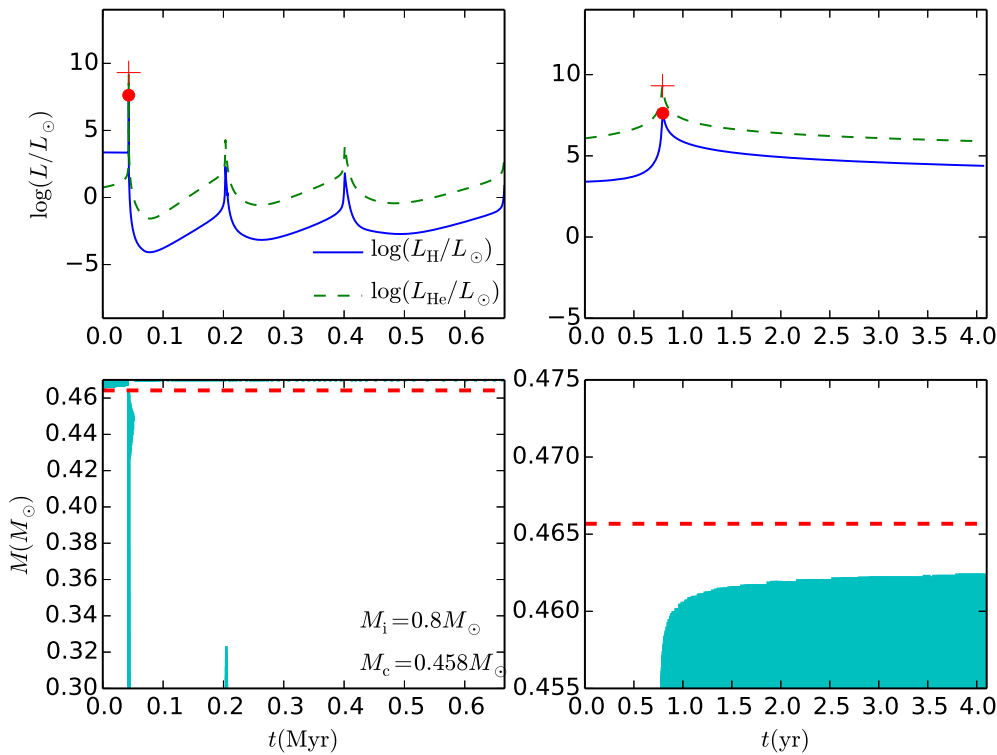}
\caption{Similar to Fig.\,\ref{fig 4}, but for $M_{\rm i}=0.8M_\odot$, $M_{\rm
c}=0.458M_\odot$. The convection developed during the He flashes never penetrates the H-rich envelope, and the strong H flash of Fig.\,\ref{fig 4} fails to appear here.}
 \label{fig 5}%
\end{figure*}

\begin{figure}
   \centering
   %%%
   \includegraphics[width=7cm,height=10cm]{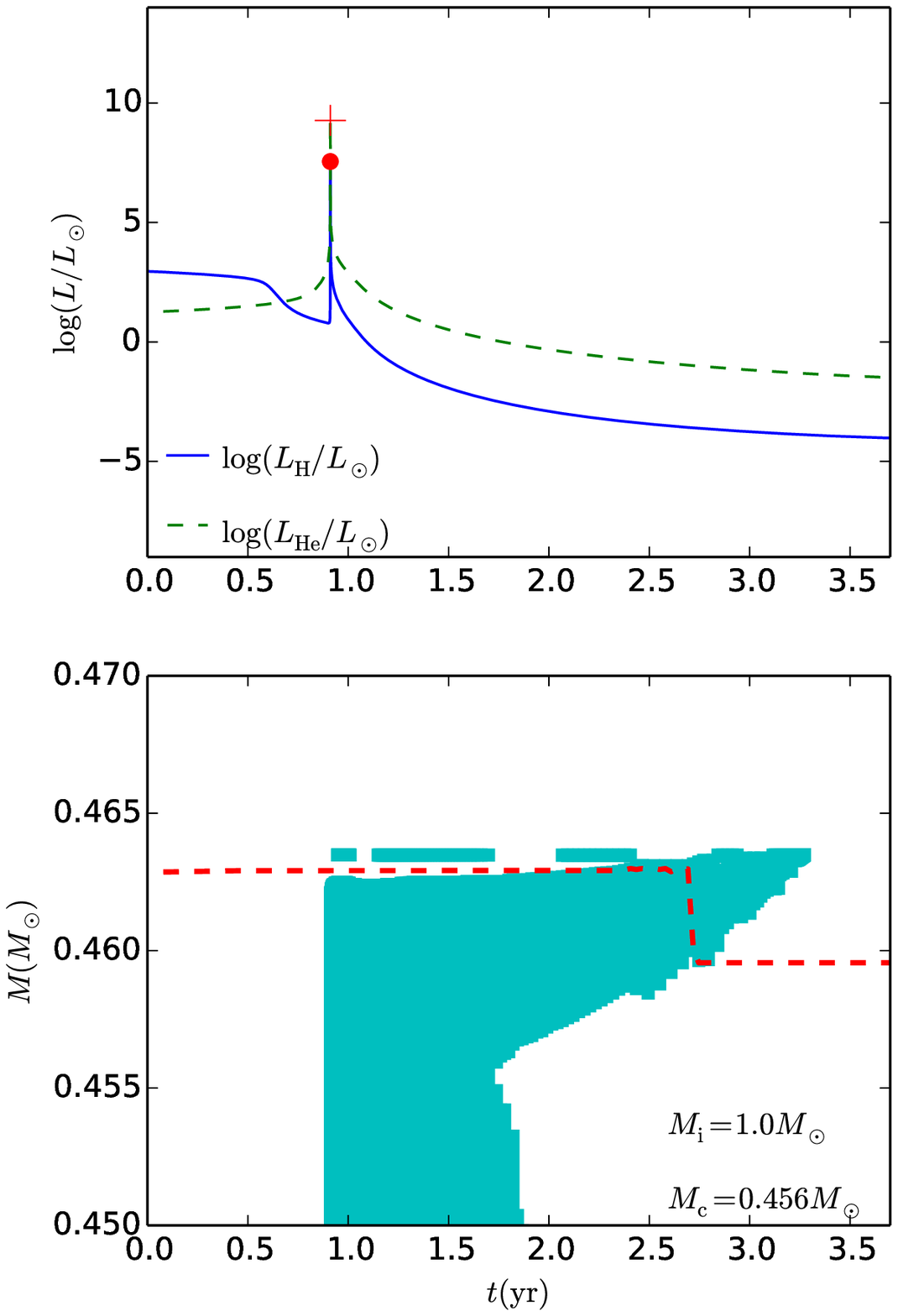}
   \caption{Similar to Fig.\,\ref{fig 4}, but for the first He and H flashes for the model of $M_{\rm i}=1.0M_\odot$, $M_{\rm c}=0.456M_\odot$ .}
   \label{fig 6}%
\end{figure}

We calculated 44 models and all the produced sdBs are presented
on the temperature-gravity ($T_{\rm eff}-{\rm log} g$) diagram (Fig.\,\ref{fig 2}),
which shows that
for each donor the produced sdB stars are clearly separated into
two
groups. One group has almost no H-rich envelope and is
crowded at the hottest temperature end of the EHB, very close to
zero-age helium main sequence (ZAHeMS, the dot-dashed lines). The
other group has a significant H-rich envelope and is spread throughout the
entire canonical EHB region. The effective temperature increases discontinuously at the transition between the two
groups, leaving a gap on $T_{\rm eff}-{\rm log} g$ diagram. Only
two products are found to be located in the gap from our models.
The products in the first group have higher He and C
abundances in the envelope than those in the second group (Table\,\ref{table 1}). In the following, we show that the convection
developed during the first He flash causes the dichotomy of
the sdB properties. For convenience, we refer to the first class as
the flash-mixing sdBs and to the second class as the canonical sdBs.

The observed sdB binaries with short-orbital periods \citep {Maxted2001,Kupfer} are
presented in Fig.\,\ref{fig 2} for comparison. Most sdBs can be
reproduced by the canonical sdBs, but several sdBs are located in
the blue extension of the canonical sdBs, that is, some are located in the gap
between the two groups and some are on the flash-mixing sdB
tracks, but the number density is obviously lower than that in the
canonical sdB region. We discuss this in
Sect. 4.1. A few sdBs are obviously bluer or hotter
 (with $T_{\rm eff}>34,000K$ and ${\rm log} g<5.6$) than the others and are  located in
a region through which our models pass quite quickly.
These samples are very probably not produced by our models because all the donors here have degenerated cores
and the produced sdBs have very similar masses of  about $0.47M_\odot$.
By comparing the location of these objects with  the locations in panel (d) of Fig. 2 of \cite{Han2002},
we found that these sdBs probably have masses more massive than $0.55M_\odot$.
Then, to produce such sdBs, the initial mass of the progenitors should be more massive than $3.2M_\odot$ according to Table 1 of \cite{Han2002}.

\subsection{Origin of the dichotomy in the sdB properties}
For each group of sdBs, we chose one typical model to show the
details during He flashes. The model for the flash-mixing sdBs has an
initial mass  of $0.8M_\odot$ and a core mass of  $0.448M_\odot$
at the onset of the CE ejection. The model for the canonical sdBs
has the same initial mass, but a core mass of $0.458M_\odot$ at the onset of CE ejection.
Figure\,\ref{fig 3} shows their evolutionary tracks after the CE ejection on the HRD
and the positions of the first He and H flashes.

The flash-mixing model is shown as the red dashed line in Fig.\,\ref{fig 3}.
For this model, the first He flash occurs when the star descends the WD cooling sequence, and it is followed
by a very strong H flash.
The He- and H-burning luminosities during the flashes are presented in Fig.\,\ref{fig 4},
where the structure of convection near the surface is also presented
to understand the surface composition of the products, then their positions on HRD.
Figure\,\ref{fig 4} shows that the convection has developed during the He flash
and penetrates the H-rich envelope (the H-rich boundary is defined as that with H
abundance $X=0.1$). Most of the H-rich matter has been  included
into the hot helium-burning region and is ignited. We therefore see an
additional strong H flash following the He flash. The maximum
luminosity of the H flash is $L_{\rm{H}}\sim10^{10}L_\odot$, even
higher than that of the He flash. As a result,  almost all H has been burnt into He}
and the remnant is almost a naked He core. All
the flash-mixing sdBs have similar processes during the first He
flash and crowd at the hottest temperature end of the EHB, very
close to ZAHeMS.

For the canonical model (the blue solid line in Fig.\,\ref{fig 3}), however, the first He
flash occurs much earlier, when the star approaches
the maximum temperature after the CE ejection, and the convection
region developed in the flash never penetrates the H-rich
envelope (see Fig.\,\ref{fig 5}). The H-rich matter is therefore not included in the
hot helium burning region, and the strong H flash that appeared in the flash-mixing
model never occurs in this model. The H-burning is moderate
in the envelope, and significant H-rich material has been left when
the stable He-core burning is established. The sdBs produced in
this way spread throughout the entire sdB region, and the exact location
on the $T_{\rm eff}-{\rm log} g$ diagram is related to the H-rich
envelope mass that is left, as studied previously
\citep{Lanz,Heber2016}.

Whether the temporary convection induced by the first He flash
penetrates the H-rich envelope depends on the position
at which the first He flash occurs, as discussed by \cite {Brown2001}. If the first He flash occurs just after the end of mass
loss (due to a strong wind, the CE ejection or Roche lobe
overflow), that is, when the star approaches the maximum
temperature or near the top of the WD cooling sequence, then
a high
entropy barrier at the bottom of the H-rich envelope prevents the
convection from penetrating the envelope.
 The entropy barrier is caused by H-shell burning
 in the red giants and is maintained when the remnant moves toward
 the maximum effective temperature just after the end of CE ejection
 as a result of relatively high H-shell burning rates during this phase,
 but it becomes lower and lower when the remnant descends the WD cooling curve because of the decrease of the energy production rate in the H-burning shell resulting from the decrease in temperature within the shell. The convection then penetrates the entropy barrier
and very easily enters the H-rich envelope.
As a consequence, most of H in the envelope is burnt
 \citep[see also][]{Iben1984,Castellani,D'Cruz,Lanz}.
This leads to the dichotomy in sdB
properties and to a discontinuous increase in $T_{\rm eff}$ between
the two classes, leaving a gap on the $T_{\rm eff}-{\rm log} g$
diagram, as presented in Fig.\,\ref{fig 2}.

The mixing between the core and the H-rich envelope may be
complete or incomplete. This furthermore divides the flash mixing into
two subtypes: deep mixing and shallow mixing. This has been shown by
\cite {Sweigart}, where the products with deep mixing are
similar to those of the flash-mixing sdBs in our study, and those
with shallow mixing are located in the gap between the
flash-mixing sdBs and the canonical sdBs. Two models in our
calculation, $M_{\rm i}=1.0M_\odot$, $M_{\rm c}=0.456M_\odot$
and $M_{\rm i}=1.5M_\odot$, $M_{\rm c}=0.458M_\odot$, are located
in the gap (see Fig.\,\ref{fig 2}). Their behaviour on the HRD is similar to
that of the canonical sdBs in Fig.\,\ref{fig 3}, except that the first He
flash occurs when the stars have entered the WD cooling
sequence but are still hot enough. The temporal evolution of the H-
and He-burning luminosity and the convection region during the He
flashes for the model of $M_{\rm i}=1.0M_\odot$, $M_{\rm
c}=0.456M_\odot$ are presented in Fig.\,\ref{fig 6}. The
convection induced by the first He flash penetrates the
H-rich envelope, but much later than in Fig.\,\ref{fig 4}, which
means that the
penetration occurs when the first He flash is extinct. In this
case, the H-rich material has not been involved in a
high-temperature region to be consumed. In the following series of
He flashes, the convection region never penetrates the H-rich
envelope again. The products with shallow mixing then have a surface
H abundance between that of deep mixing and that of the canonical
sdBs, and remain in the gap between the two classes.

\section{Discussions}

\subsection{Modelling the CE ejection process}
\begin{figure*}
\centering
   %%%
\includegraphics[width=14cm]{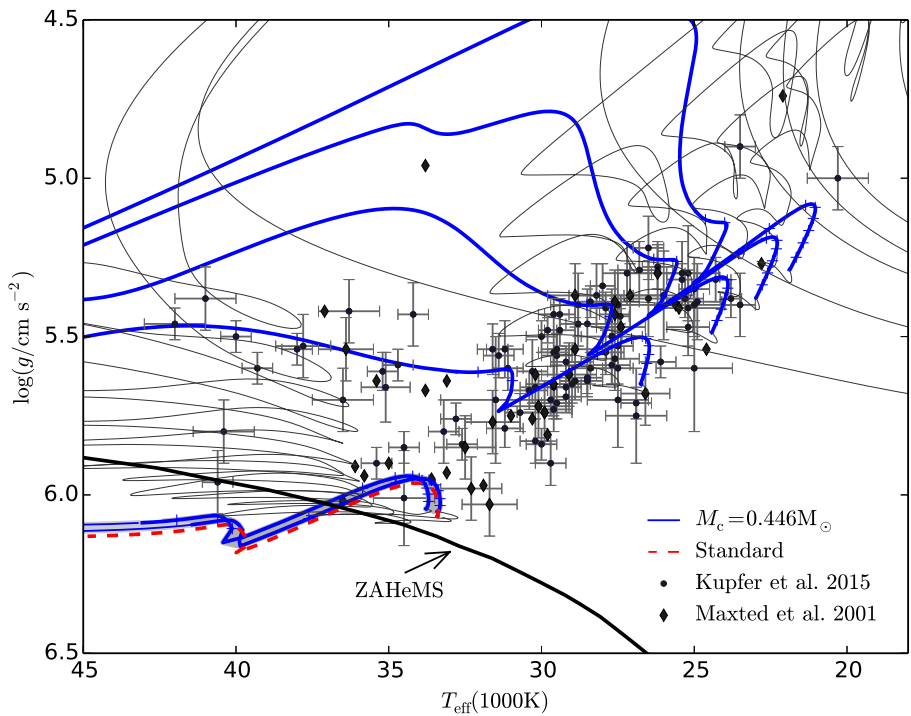}
\caption{Evolutionary tracks of the produced sdB stars from the
star with an initial mass of $1.0M_\odot$ and a core mass of
$0.446M_\odot$. The sdB mass, $M_{\rm sdB}$, is artificially set
to be equal to 0.452, 0.454,...,0.472 and $0.474M_\odot$ (in steps
of $0.002M_\odot$). The products crowd at the hottest end of EHB
for $M_{\rm sdB}=0.452-0.466M_\odot$
(located in the shaded region between the
lowest two dashed lines), and spread throughout the canonical sdB region
when $M_{\rm sdB}\ge 0.468M_\odot$ (from bottom to top, $M_{\rm
sdB}=0.468,0.470,0.472,$ and $0.474M_\odot$). The red
solid line shows the standard model ($M_{\rm sdB}=0.453M_\odot$). SdBs
with short
orbital periods from  \cite{Maxted2001,Kupfer} are presented in the figure for
comparison. The grey lines are the tracks during helium flashes
before stable He-core burning.}
              \label{fig 7}%
\end{figure*}

We mimicked the CE ejection process using a very high
mass-loss rate ($10^{-3}M_\odot{\rm yr^{-1}}$) and stopped the
ejection process when the star began to collapse. In other words, the radius
of the star, $R$, is equal to the initial radius at the beginning
of mass loss, $R_{\rm 0}$. Although this modelling is physically
reasonable, there is still a large uncertainty because of the lack of
detailed dynamical simulations. The mass of the envelope, which is
determined by the endpoint of the ejection process, is a key
factor that can alter the properties of the produced sdBs, since
it is related to the development of the convection region during the
He flashes. The end of the CE ejection process is therefore very important here. In this subsection, we
examine this by artificially fixing the remnant mass after
the CE ejection process. To do this, the ejection process is stopped when
the star has a mass equal to a specified value. The examined star
has an initial mass of $M_{\rm i}=1.0M_\odot$ and a core mass of
$M_{\rm c}=0.446M_\odot$ at the onset of the CE ejection. The
produced sdB is a flash-mixing sdB with a mass of $0.453M_\odot$
from our study in Sect. 3 (the standard model hereafter).
Various stellar masses after the CE ejection process are adopted: $M_{\rm sdB}=0.452-0.472M_\odot$ in steps of $0.002M_\odot$.
The results are shown in Fig.\,\ref{fig 7}, where the products crowd at the
hottest end of EHB when $M_{\rm sdB}=0.452-0.466M_\odot$, and
spread throughout the canonical sdB region when $M_{\rm sdB}\ge
0.468M_\odot$. This means that the envelope mass (or the
position at which the CE ejection terminates) determines the
place of the produced sdBs on the $T_{\rm eff}-{\rm log}g$
diagram. Our study shows that for a given initial stellar mass
and a given core mass at the onset of the CE, the star has a more massive  envelope when the CE ejection
stops earlier. The product is then more likely to be a
canonical sdB, and the opposite holds when the CE ejection stops
later.

Very few sdB binaries with short orbital periods are
located on or near the flashing-mixing sdB tracks, as described in
Sect. 2. This suggests that the envelope mass after the CE ejection is
probably more massive than that expected before. We
here stopped the ejection when the donor contracted back into its Roche lobe, $R_{\rm L}$.
The value of $R_{\rm L}$ is $\approx R_{\rm 0}$ at the onset of the CE and decreases with the
ejection process. The assumption that the ejection ends as
$R<R_{\rm 0}$ is therefore an upper limit and gives the maximum
envelope mass left on the remnant. For example, we could have obtained a lower envelope mass if we had replaced $R_{\rm L}$ with the critical Roche-lobe radius of the remnant or the core radius before the CE ejection. Furthermore, \cite{Hall2014} showed that the post-CE stripped remnant radius is smaller than that of the pre-CE core when the star is in HG or on RGB (Table 3 of that paper). This allows the companions to spiral in more closely to the core than before, and the envelope mass of the remnant decreases further if this is the case. A falling-back process, similar
to the process that occurs in nova, probably occurs in the CE ejection and
provides an appropriate way of increasing the envelope mass. The
detailed process is beyond of the scope of this paper.

\subsection{Mass loss during flashes}
We did not include any mass-loss process except
for the CE ejection for simplicity. However, the star obviously
loses some material when it evolves on the RGB or during strong
flashes. The stellar wind on the RGB may change some orbit parameters
of sdB binaries such as the orbital period, but it probably does
not affect
the properties of the sdB stars, which are determined by the evolution after the envelope is stripped. The mass loss
during the flashes will reduce the envelope mass of a pre-sdB
star, then alters its final position on the $T_{\rm eff}-{\rm
log}g$ diagram when stable He-core burning begins. There are two
ways of mass loss during the flashes: a strong stellar wind driven
by the violent flashes, and Roche lobe overflow when the pre-sdB
star expands to exceed its Roche lobe radius. The effect of mass
loss on the sdB mass is negligible since the mass loss is too
small in comparison to $M_{\rm sdB}$, but the effect on the
envelope mass may be significant since the envelope mass is itself
very low. The mass loss is expected to be higher for the
flash-mixing sdBs than for the canonical sdBs because of the
additional strong H flash after the first He flash. For the
flash-mixing sdBs, the surface He and C abundance presented in
Table\,\ref{table 1} may be further enhanced, but their locations on the
$T_{\rm eff}-{\rm log}g$ diagram will not change much since these
objects have nearly H-exhausted envelope. For the canonical
sdBs, the mass loss reduces the envelope mass, and the tracks will
move towards the lower left on the $T_{\rm eff}-{\rm log}g$
diagram.  As a consequence, the sdBs immediately above the gap may change into
flash-mixing or the shallow-mixing sdBs, while the upper right lines,
where very few observational samples are located, will move downwards,
which is better consistent with the observations.

\subsection{Effect of metallicity}

\begin{figure*}
\centering
   %%%
\includegraphics[width=14cm]{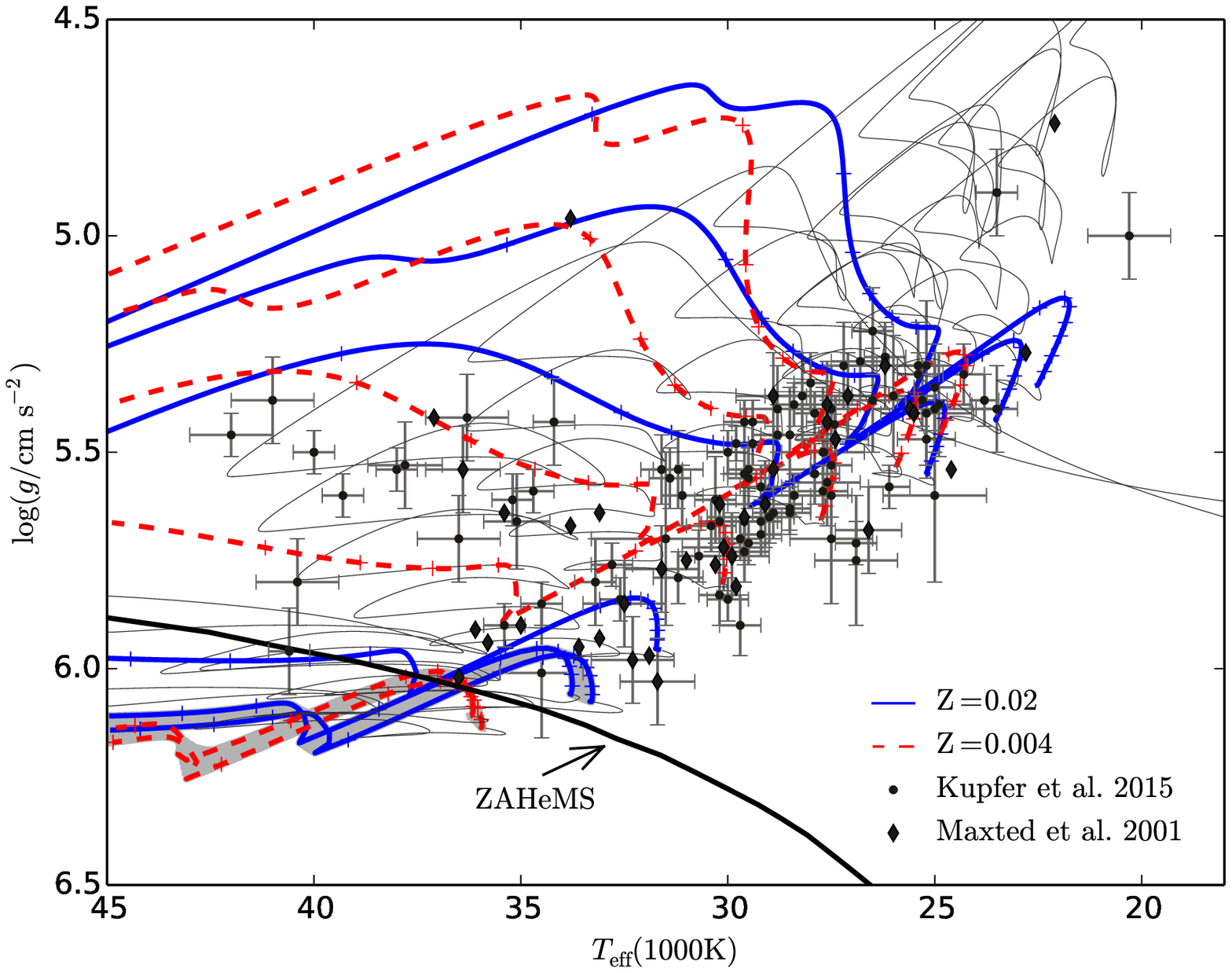}
\caption{Evolutionary tracks of the produced sdB stars on the
$T_{\rm eff}-{\rm log} g$ diagram. The star has an initial mass of
$1M_\odot$ with a metallicity of $Z=0.004$ (the dashed lines). The
products crowd at the hottest end of EHB for the core mass $M_{\rm
c}=0.453-0.456M_\odot$ (located in the shaded narrow region of the lowest
two lines), and spread throughout the canonical sdB region when $M_{\rm
c}\ge 0.466M_\odot$ (from bottom to top, $M_{\rm
c}=0.466,0.468,0.470,$ and $0.472M_\odot$). The solid
lines show the results of $Z=0.02$ for the star with the same
initial mass. The age difference between adjacent crosses is $10^7
{\rm yr}$.  SdBs with short orbital periods from  \cite{Maxted2001,Kupfer} are
presented in the figure for comparison. The grey lines are the
tracks during helium flashes before stable He-core burning.
\label{fig 8}}%
\end{figure*}

All the studies above are for Population I stars ($Z=0.02$). Here
we show some models for $Z=0.004$ to demonstrate the effect of
metallicity on the final results. The star has an initial mass of
$1.0M_\odot$ and various core masses at the onset of the CE
process (see Table\,\ref{table 2}). The results are shown in Fig.\,\ref{fig 8}. They are
very similar to those of Population I stars except for the exact positions
on the $T_{\rm eff}-{\rm log}g$ diagram, that is, the produced
sdBs appear as two groups, one group crowds at the hottest end of
EHB (for $M_{\rm c}=0.453-0.456M_\odot$), and the other group spreads
throughout the canonical sdB region (for $M_{\rm c}\ge 0.466M_\odot$).
The whole tracks of $Z=0.004$ move towards the lower left on the $T_{\rm
eff}-{\rm log}g$ diagram, that is, sdBs with low initial metallicities
have higher surface gravities (or smaller radii) and higher effective temperatures
because of the lower opacity of the envelope during He-core burning.
The luminosity of sdBs is mainly determined by the core mass during He-core burning.
For $Z=0.004$, the minimum core mass for He ignition at the first He flash is $0.460M_\odot$ (Table\,\ref{table 2}),
but it is $0.449M_\odot$ for $Z=0.02$ (Table\,\ref{table 1}).
The flash-mixing sdB stars with $Z=0.004$ are therefore more luminous than those with $Z=0.02$.
The study here suggests that the observed sdBs in the gap between the flash-mixing sdBs and the canonical
sdBs can be well understood by adopting various metallicities.

\subsection{He-rich sdB stars}
\begin{figure*}
\centering
\includegraphics[width=14cm]{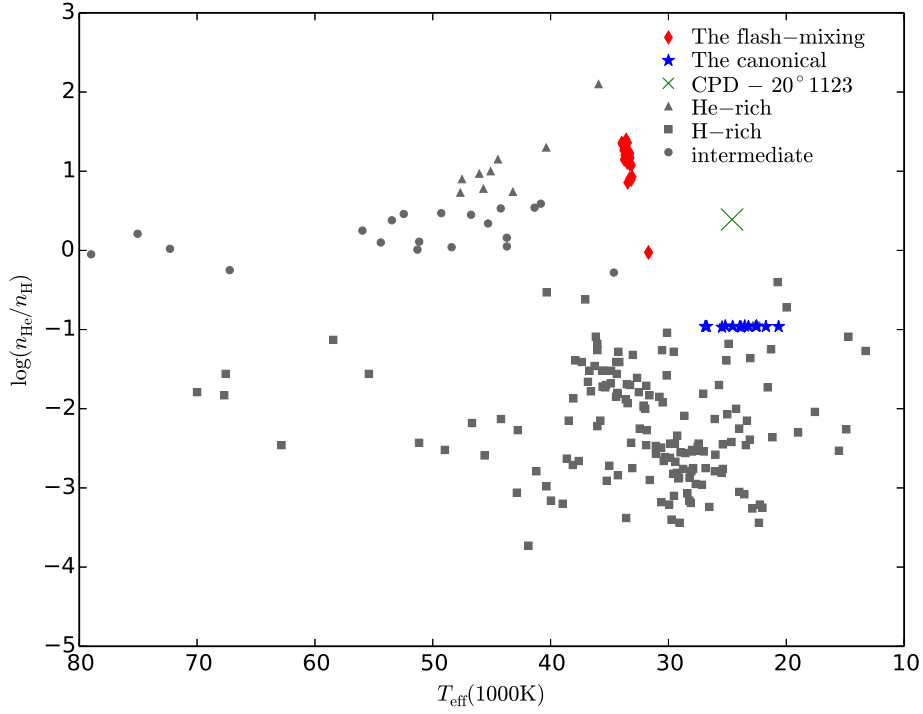}
\caption{Ratio of He to H abundance at the surface related to
the solar value, [He/H], versus effective temperature, for our
theoretical models and the observations. The sdB samples are from
\cite{Peter} and have been divided into three
types, i.e. H-rich, intermediate He-rich, and He-rich, as indicated
in the figure. The diamonds are for the flash-mixing sdBs and the
stars are for the canonical sdBs, where the He and H abundances
are chosen at the onset of stable helium-core burning. The cross
stands for CPD-$20^{\rm o}1123$, an intermediate He-rich
sdB with an orbital period of 2.3698 d \citep{Naslim}.}
\label{fig 9}%
\end{figure*}

\begin{figure*}
\centering
\includegraphics[width=14cm]{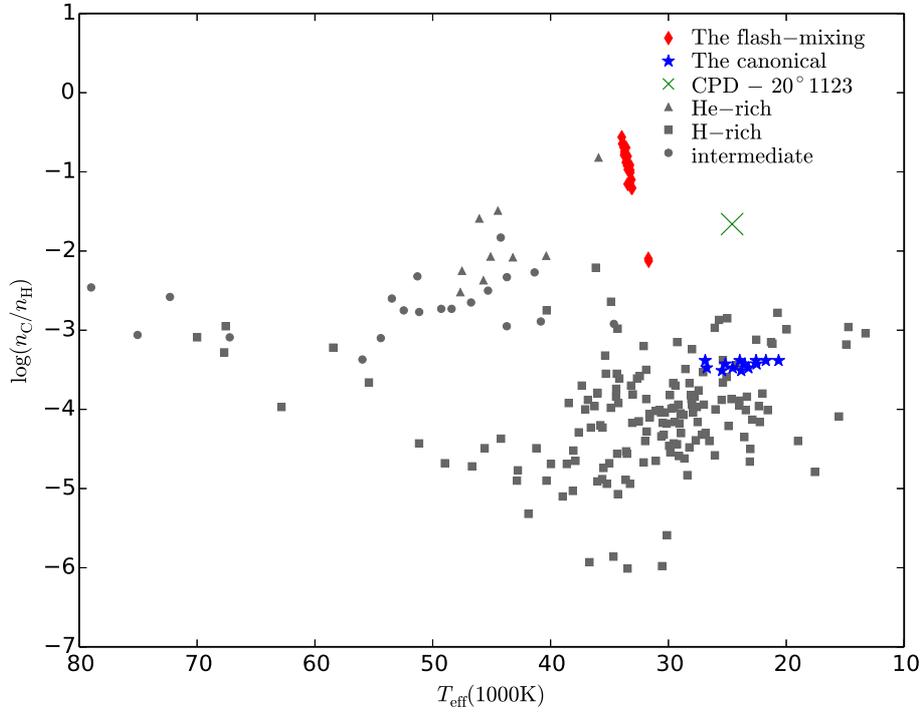}
\caption{Similar to Fig.\,\ref{fig 9}, but for [C/H].}
\label{fig 10}%
\end{figure*}

Since He and C have been enhanced in the envelope during the flash
mixing (see Table\,\ref{table 1}), the flash-mixing sdBs may be connected with
the He-rich sdB stars. Figures\,\ref{fig 9} and\,\ref{fig 10} show comparisons of our
models with observations in the $T_{\rm eff}-[{\rm He/H}]$
diagram and in the $T_{\rm eff}-[{\rm C/H}]$ diagram. The observed atmospheric parameters of sdBs are from
\cite{Peter}, who divided the sample into
three types according to the ratio of He to H abundance, $\log{(n_{\rm He}/n_{\rm H})}$: H-rich if $\log{(n_{\rm He}/n_{\rm H})}<-0.349$,
intermediate He-rich if $-0.349\le\log{(n_{\rm He}/n_{\rm H})}\le0.6,$ and He-rich if
$\log{(n_{\rm He}/n_{\rm H})}>0.6$. Figure\,\ref{fig 9} shows that the flash-mixing sdBs
(the diamonds) have similar surface abundances as
intermediate He-rich and He-rich sdB stars, while the canonical
sdBs (the stars) are located near the boundary that separates the
H-rich and intermediate He-rich sdBs. We note that with the evolving
sdBs, the effective temperature increases and the surface He
abundance probably reduces when the gravity setting is considered.
The theoretical models then move left downwards, that is, the
flash-mixing models may cross the intermediate He-rich and
He-rich samples and the canonical models will cross the H-rich
samples.

\begin{figure*}
\centering
\includegraphics[width=14cm]{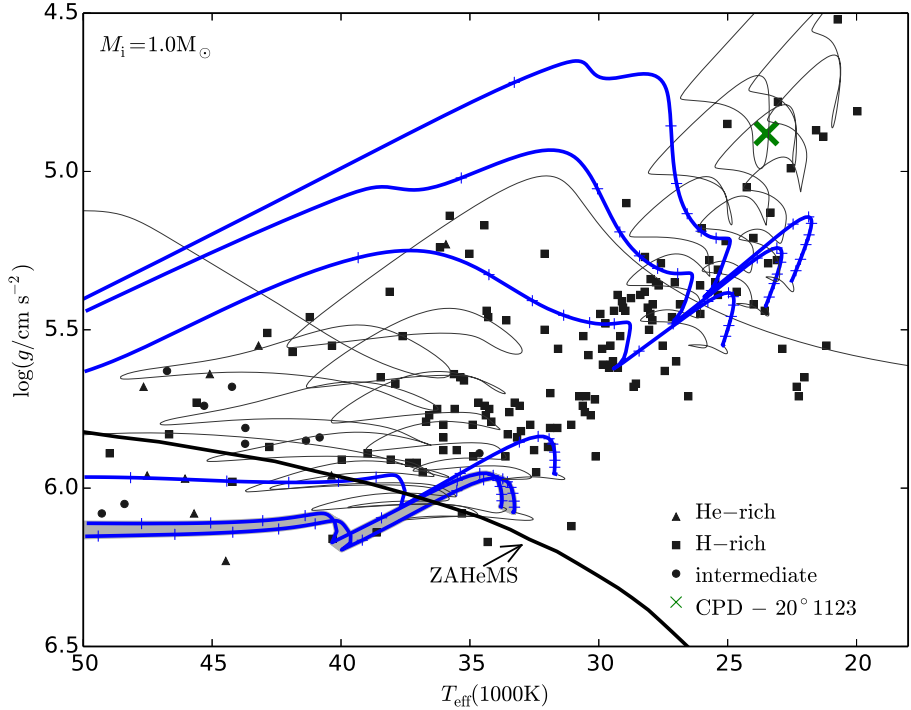}
\caption{Evolutionary tracks of the produced sdBs and the sdB
samples with atmospheric He abundance \citep{Peter} on the effective temperature-gravity diagram.
The shaded area between the lowest two lines is
the region of the evolutionary tracks that are occupied by the flash-mixing models.
The age difference between adjacent crosses on the evolutionary
tracks is $10^7 {\rm yr}$. The cross stands for
CPD-$20^{\rm o}1123$, an intermediate He-rich sdB with an orbital
period of 2.3698 d \citep{Naslim}. The grey
lines are the tracks during helium flashes before stable He-core
burning.}
\label{fig 11}%
\end{figure*}

Figure \ref{fig 11} presents sdB samples on the $T_{\rm eff}-{\rm
log}g$ diagram compared with theoretical evolutionary
tracks. The figure shows that the intermediate He-rich and He-rich
sdBs are indeed spread around the evolutionary tracks of
flash-mixing models and also indicates that most of the
(intermediate) He-rich stars can be well understood by the flash
mixing  \citep[see also][]{Miller}. In particular,
the flash-mixing sdBs that originated from the CE ejection channel are
on short-orbital periods and are probably responsible for
(intermediate) He-rich sdBs with short orbital periods. However,
only one intermediate He-rich sdB, CPD-$20^{\rm o}1123$, has been
found on a short orbital period \citep {Naslim}, and its location
on the $T_{\rm eff}-{\rm log}g$ diagram seems not relevant to the
flash-mixing model. The lack of He-rich sdBs on short-orbital
periods further indicates that the flash mixing does probably
not occur in the products of the CE ejection. The reason for
this probably is the relatively
massive envelope mass, which in turn resulted from a falling-back process after
the CE ejection, as discussed in Sect. 4.1.

\subsection{Blue hook stars in GCs}
Blue hook stars occupy a very blue position on the HB, but with
a fainter luminosity than normal EHB stars. Their formation
mechanism is not very clear. \cite {Brown2001}
suggested that the late hot He flash, which occurs as the star
descends the WD cooling curve, can reproduce such objects. The progenitors in the study of Brown and colleagues undergo unusually huge mass loss on
the RGB, and the products are constructed based on some assumptions
that are due to the numerical difficulties in dealing with the He flashes. \cite {Lei} reproduced the blue hook stars in NGC 2808 by employing tidally
enhanced stellar wind in binaries. The
evolutions after the mass loss are very similar to the evolution shown in
this paper since the remnants have very similar structures
regardless of how the the mass stripping occurs (CE ejection or
strong stellar wind). Our study then provides a new way to forming
the blue hook stars in GCs. Furthermore, sdB stars from stable
mass transfer may also have similar processes if the envelope mass
is low enough. The flash mixing is crucial in all of the
suggested formation channels and leads to similar observational
properties for the blue hook stars themselves. The hints of their
origins probably exist in the orbital parameters, such as the orbital
period if they were in binaries.

\subsection{Comparison with sdBs with long orbital periods}
 Using MESA, \cite{Vos2015}  and \cite{Lei,Lei2016} obtained
several sdBs with long orbital periods through stable mass transfer or tidally enhanced stellar wind.
These sdBs  with long orbital periods are also divided into three classes: flash-mixing sdBs, shallow-mixing sdBs, and canonical sdBs. However, we did not find any discussions on the numbers of the shallow-mixing sdBs. On the other hand, we found relatively broad orbital period ranges for sdBs with late flashes (to produce the flash-mixing and the shallow mixing sdBs) in \cite{Lei2016}. Based on our results, which are that the envelope mass of the pre-sdBs determines the properties of the sdB stars and that
the envelope mass range for the shallow-mixing sdBs is very narrow, most of the products from the models in the orbital period ranges for late flashes may be the flash-mixing sdBs and not the shallow-mixing sdBs. If a large portion of the produced sdBs in the orbital period ranges for late flashes are shallow mixing, the tidally enhanced wind is then very favourable for forming sdBs with an
envelope mass exactly in the range for the shallow-mixing sdBs.

\section{Conclusions}
Employing the code MESA, we studied the properties of sdBs produced from
the CE channel. The stars appear as two distinct groups on the
$T_{\rm{eff}}-\log{g}$ diagram: the flash-mixing sdBs, and the
canonical sdBs, as defined in this paper. The flash-mixing sdBs have almost no H-rich envelope and are crowded at the hottest
temperature end of EHB, while the canonical sdBs have a significant
H-rich envelope and are spread throughout the entire canonical EHB region.

The key factor for the dichotomy of the sdB properties is the
extent of convection during the first helium flash. For the
flash-mixing sdBs, the star enters the CE process earlier and
has a lower He core mass. The first helium flash occurs when the
star descends the WD cooling curve and is followed by a
violent hydrogen flash, which is triggered by convective element
mixing between H-rich envelope and He-burning region. Hydrogen in
the envelope is then almost exhausted and the products are almost
naked He cores. For the canonical sdBs, the star enters the
CE process later and has a more massive He core. The first helium
flash occurs much earlier, and the convection induced by the
helium flash never penetrates the H-rich envelope because of
the
high entropy barrier of H-rich envelope. The products then remain
significant H-rich envelope and are spread throughout the entire canonical
EHB region. Their positions are determined by the H-rich envelope
mass as studied in many previous works. Therefore, the dichotomy of
the sdB properties from the CE ejection channel is intrinsic and
caused by the interior structure of the star after the CE
ejection.

The treatment of convection and the modelling of the CE ejection
process will greatly change the parameter spaces for the two typical
groups of sdB stars. For a given initial stellar mass and a given
core mass at the onset of the CE, the star will be more massive and have a more-massive
envelope if the CE ejection stops earlier. The produced sdB is more likely to be a canonical sdB, and
the opposite holds for later ejection. The fact that very few short-orbital period sdB
binaries are located in the flash-mixing sdB region means that the
sdBs produced from the CE ejection have a more massive envelope mass
than expected. The lack of He-rich sdBs with short orbital periods also
indicates that the flash mixing is rare in the products of the CE
ejection. A falling-back process after the CE ejection, similar to
what occurs in nova, is an appropriate way of increasing the
envelope mass, then prevents the flash mixing.

\begin{table*}
\caption{Models investigated for a metallicity of $Z=0.004$. The
star has an initial mass of $1.0M_\odot$. The first three columns
show the core mass at the onset of the CE process $M_{\rm c}$, the core mass just before the first He flash $M_{\rm c}^*$, and
the remnant mass after the CE process $M_{\rm sdB}$, respectively.
Columns 4-8 are the total hydrogen mass $M_{\rm H}$, surface
He abundance ($Y$) , surface C abundance ($X_{\rm C}$),  surface N abundance ($X_{\rm N}$), and surface
O abundance ($X_{\rm O}$), respectively, as He burns stably in the core.
The masses are in units of solar mass.}             % title of Table
\label{table 2}      % is used to refer this table in the text
\centering                          % used for centering table
\begin{tabular}{cccccccccc}        % centered columns (4 columns)
\hline\hline                 % inserts double horizontal lines
$M_{\rm c}$&$M_{\rm c}^*$&$M_{\rm sdB}$&$M_{\rm H}$&Y&$X_{\rm C}$&$X_{\rm N}$&$X_{\rm O}$\\
\hline          % inserts single horizontal line
  0.453  &0.460 &0.460  &$0.820\times10^{-4}$&0.964 &0.015&0.013&$1.118\times10^{-4}$\\
  0.454  &0.460 &0.461  &$0.916\times10^{-4}$&0.956 &0.011&0.017&$1.175\times10^{-4}$\\
  0.456  &0.462 &0.463  &$1.178\times10^{-4}$&0.959 &0.015&0.015&$1.187\times10^{-4}$\\
  0.458  &0.464 &0.465  &$2.200\times10^{-4}$&0.953 &0.014&0.015&$9.938\times10^{-5}$\\
  0.46  &0.466  &0.467  &$1.304\times10^{-4}$&0.956 &0.017&0.014&$1.152\times10^{-4}$\\
  0.462 &0.469  &0.469  &$0.926\times10^{-4}$&0.952 &0.020&0.012&$1.225\times10^{-4}$\\
  0.464  &0.470 &0.471  &$2.329\times10^{-4}$&0.944 &0.023&0.011&$1.254\times10^{-4}$\\
  0.466  &0.472 &0.473  &$5.177\times10^{-4}$&0.269 &0.001&$3.113\times10^{-4}$&$1.871\times10^{-3}$\\
  0.468  &0.473 &0.475  &$1.719\times10^{-3}$&0.269 &0.001&$3.113\times10^{-4}$&$1.871\times10^{-3}$\\
  0.47  &0.473  &0.478  &$3.395\times10^{-3}$&0.269 &0.001&$3.113\times10^{-4}$&$1.871\times10^{-3}$\\
  0.472  &0.473 &0.480  &$4.759\times10^{-3}$&0.269 &0.001&$3.113\times10^{-4}$&$1.871\times10^{-3}$\\
\hline\hline
%inserts single line
\end{tabular}
\label{table 2}
\end{table*}

The effective temperature at
the transition between the two groups decreases discontinuously, leaving a gap on the
$T_{\rm{eff}}-\log{g}$ diagram. Only two products are found to be
in the gap from our model grid. The properties of the two models
are consistent with the properties of shallow mixing (mixing is incomplete)
and can well explain the blue hook stars in NGC 2808. Various
formation scenarios for blue hook stars give similar
characteristics of these objects, and hints of their origins
probably exist in the orbital parameters if they were in binaries.


\begin{acknowledgements}
This work is supported by the Natural Science Foundation of China
(Nos. 11422324, 11521303, 11390374), by Yunnan province
(Nos. 2012HB037, 2013HA005) and by the Chinese Academy of Sciences
(No. KJZD-EW-M06-01).
\end{acknowledgements}

\bibliographystyle{aa-note}%

\bibliography{bibfile}

\begin{thebibliography}{42}
\expandafter\ifx\csname natexlab\endcsname\relax\def\natexlab#1{#1}\fi

\bibitem[{{Altmann} {et~al.}(2004){Altmann}, {Edelmann}, \& {de
  Boer}}]{Altmann}
{Altmann}, M., {Edelmann}, H., \& {de Boer}, K.~S. 2004, \aap, 414, 181 \csname
  Altmannlink\endcsname~\csname Altmannnote\endcsname

\bibitem[{{Barlow} {et~al.}(2013){Barlow}, {Wade}, \& {Liss}}]{Barlow}
{Barlow}, B., {Wade}, R.~A., \& {Liss}, S. 2013, in American Astronomical
  Society Meeting Abstracts, Vol. 221, American Astronomical Society Meeting
  Abstracts \#221, 142.17 \csname Barlowlink\endcsname~\csname
  Barlownote\endcsname

\bibitem[{{Brown} {et~al.}(2000){Brown}, {Bowers}, {Kimble}, {Sweigart}, \&
  {Ferguson}}]{Brown2000}
{Brown}, T.~M., {Bowers}, C.~W., {Kimble}, R.~A., {Sweigart}, A.~V., \&
  {Ferguson}, H.~C. 2000, \apj, 532, 308 \csname
  Brown2000link\endcsname~\csname Brown2000note\endcsname

\bibitem[{{Brown} {et~al.}(2001){Brown}, {Sweigart}, {Lanz}, {Landsman}, \&
  {Hubeny}}]{Brown2001}
{Brown}, T.~M., {Sweigart}, A.~V., {Lanz}, T., {Landsman}, W.~B., \& {Hubeny},
  I. 2001, \apj, 562, 368 \csname Brown2001link\endcsname~\csname
  Brown2001note\endcsname

\bibitem[{{Castellani} \& {Castellani}(1993)}]{Castellani}
{Castellani}, M. \& {Castellani}, V. 1993, \apj, 407, 649 \csname
  Castellanilink\endcsname~\csname Castellaninote\endcsname

\bibitem[{{Charpinet} {et~al.}(2010){Charpinet}, {Green}, {Baglin}, {Van
  Grootel}, {Fontaine}, {Vauclair}, {Chaintreuil}, {Weiss}, {Michel},
  {Auvergne}, {Catala}, {Samadi}, \& {Baudin}}]{Charpinet}
{Charpinet}, S., {Green}, E.~M., {Baglin}, A., {et~al.} 2010, \aap, 516, L6
  \csname Charpinetlink\endcsname~\csname Charpinetnote\endcsname

\bibitem[{{Chen} {et~al.}(2013){Chen}, {Han}, {Deca}, \&
  {Podsiadlowski}}]{Chen2013}
{Chen}, X., {Han}, Z., {Deca}, J., \& {Podsiadlowski}, P. 2013, \mnras, 434,
  186 \csname Chen2013link\endcsname~\csname Chen2013note\endcsname

\bibitem[{{Copperwheat} {et~al.}(2011){Copperwheat}, {Morales-Rueda}, {Marsh},
  {Maxted}, \& {Heber}}]{Copperwheat}
{Copperwheat}, C.~M., {Morales-Rueda}, L., {Marsh}, T.~R., {Maxted}, P.~F.~L.,
  \& {Heber}, U. 2011, \mnras, 415, 1381 \csname
  Copperwheatlink\endcsname~\csname Copperwheatnote\endcsname

\bibitem[{{D'Cruz} {et~al.}(1996){D'Cruz}, {Dorman}, {Rood}, \&
  {O'Connell}}]{D'Cruz}
{D'Cruz}, N.~L., {Dorman}, B., {Rood}, R.~T., \& {O'Connell}, R.~W. 1996, \apj,
  466, 359 \csname D'Cruzlink\endcsname~\csname D'Cruznote\endcsname

\bibitem[{{Ferguson} {et~al.}(1991){Ferguson}, {Davidsen}, {Kriss}, {Blair},
  {Bowers}, {Dixon}, {Durrance}, {Feldman}, {Henry}, {Kruk}, {Moos}, {Vancura},
  {Long}, \& {Kimble}}]{Ferguson}
{Ferguson}, H.~C., {Davidsen}, A.~F., {Kriss}, G.~A., {et~al.} 1991, \apjl,
  382, L69 \csname Fergusonlink\endcsname~\csname Fergusonnote\endcsname

\bibitem[{{Hall} \& {Tout}(2014)}]{Hall2014}
{Hall}, P.~D. \& {Tout}, C.~A. 2014, \mnras, 444, 3209 \csname
  Hall2014link\endcsname~\csname Hall2014note\endcsname

\bibitem[{{Han}(2008)}]{Han2008}
{Han}, Z. 2008, \aap, 484, L31 \csname Han2008link\endcsname~\csname
  Han2008note\endcsname

\bibitem[{{Han} {et~al.}(2007){Han}, {Podsiadlowski}, \&
  {Lynas-Gray}}]{Han2007}
{Han}, Z., {Podsiadlowski}, P., \& {Lynas-Gray}, A.~E. 2007, \mnras, 380, 1098
  \csname Han2007link\endcsname~\csname Han2007note\endcsname

\bibitem[{{Han} {et~al.}(2003){Han}, {Podsiadlowski}, {Maxted}, \&
  {Marsh}}]{Han2003}
{Han}, Z., {Podsiadlowski}, P., {Maxted}, P.~F.~L., \& {Marsh}, T.~R. 2003,
  \mnras, 341, 669 \csname Han2003link\endcsname~\csname Han2003note\endcsname

\bibitem[{{Han} {et~al.}(2002){Han}, {Podsiadlowski}, {Maxted}, {Marsh}, \&
  {Ivanova}}]{Han2002}
{Han}, Z., {Podsiadlowski}, P., {Maxted}, P.~F.~L., {Marsh}, T.~R., \&
  {Ivanova}, N. 2002, \mnras, 336, 449 \csname Han2002link\endcsname~\csname
  Han2002note\endcsname

\bibitem[{{Heber}(2009)}]{Heber2009}
{Heber}, U. 2009, \araa, 47, 211 \csname Heber2009link\endcsname~\csname
  Heber2009note\endcsname

\bibitem[{{Heber}(2016)}]{Heber2016}
{Heber}, U. 2016, \pasp, 128, 082001 \csname Heber2016link\endcsname~\csname
  Heber2016note\endcsname

\bibitem[{{Iben}(1984)}]{Iben1984}
{Iben}, Jr., I. 1984, \apj, 277, 333 \csname Iben1984link\endcsname~\csname
  Iben1984note\endcsname

\bibitem[{{Ivanova} {et~al.}(2013){Ivanova}, {Justham}, {Chen}, {De Marco},
  {Fryer}, {Gaburov}, {Ge}, {Glebbeek}, {Han}, {Li}, {Lu}, {Marsh},
  {Podsiadlowski}, {Potter}, {Soker}, {Taam}, {Tauris}, {van den Heuvel}, \&
  {Webbink}}]{Ivanova}
{Ivanova}, N., {Justham}, S., {Chen}, X., {et~al.} 2013, \aapr, 21, 59 \csname
  Ivanovalink\endcsname~\csname Ivanovanote\endcsname

\bibitem[{{Jeffery} \& {Pollacco}(1998)}]{Jeffery}
{Jeffery}, C.~S. \& {Pollacco}, D.~L. 1998, \mnras, 298, 179 \csname
  Jefferylink\endcsname~\csname Jefferynote\endcsname

\bibitem[{{Koen} {et~al.}(1998){Koen}, {Orosz}, \& {Wade}}]{Koen}
{Koen}, C., {Orosz}, J.~A., \& {Wade}, R.~A. 1998, \mnras, 300, 695 \csname
  Koenlink\endcsname~\csname Koennote\endcsname

\bibitem[{{Kupfer} {et~al.}(2015){Kupfer}, {Geier}, {Heber}, {Ostensen},
  {Barlow}, {Maxted}, {Heuser}, {Schaffenroth}, \& {Gaensicke}}]{Kupfer}
{Kupfer}, T., {Geier}, S., {Heber}, U., {et~al.} 2015, VizieR Online Data
  Catalog, 357 \csname Kupferlink\endcsname~\csname Kupfernote\endcsname

\bibitem[{{Lanz} {et~al.}(2004){Lanz}, {Brown}, {Sweigart}, {Hubeny}, \&
  {Landsman}}]{Lanz}
{Lanz}, T., {Brown}, T.~M., {Sweigart}, A.~V., {Hubeny}, I., \& {Landsman},
  W.~B. 2004, \apj, 602, 342 \csname Lanzlink\endcsname~\csname
  Lanznote\endcsname

\bibitem[{{Lei} {et~al.}(2015){Lei}, {Chen}, {Zhang}, \& {Han}}]{Lei}
{Lei}, Z., {Chen}, X., {Zhang}, F., \& {Han}, Z. 2015, \mnras, 449, 2741
  \csname Leilink\endcsname~\csname Leinote\endcsname

\bibitem[{{Lei} {et~al.}(2016){Lei}, {Zhao}, {Zeng}, {Shen}, {Lan}, {Jiang}, \&
  {Han}}]{Lei2016}
{Lei}, Z., {Zhao}, G., {Zeng}, A., {et~al.} 2016, \mnras, 463, 3449 \csname
  Lei2016link\endcsname~\csname Lei2016note\endcsname

\bibitem[{{Maxted} {et~al.}(2001){Maxted}, {Heber}, {Marsh}, \&
  {North}}]{Maxted2001}
{Maxted}, P.~F.~L., {Heber}, U., {Marsh}, T.~R., \& {North}, R.~C. 2001,
  \mnras, 326, 1391 \csname Maxted2001link\endcsname~\csname
  Maxted2001note\endcsname

\bibitem[{{Maxted} {et~al.}(2000{\natexlab{a}}){Maxted}, {Marsh}, \&
  {North}}]{Maxted2000a}
{Maxted}, P.~F.~L., {Marsh}, T.~R., \& {North}, R.~C. 2000{\natexlab{a}},
  \mnras, 317, L41 \csname Maxted2000alink\endcsname~\csname
  Maxted2000anote\endcsname

\bibitem[{{Maxted} {et~al.}(2000{\natexlab{b}}){Maxted}, {Moran}, {Marsh}, \&
  {Gatti}}]{Maxted2000b}
{Maxted}, P.~F.~L., {Moran}, C.~K.~J., {Marsh}, T.~R., \& {Gatti}, A.~A.
  2000{\natexlab{b}}, \mnras, 311, 877 \csname
  Maxted2000blink\endcsname~\csname Maxted2000bnote\endcsname

\bibitem[{{Miller Bertolami} {et~al.}(2008){Miller Bertolami}, {Althaus},
  {Unglaub}, \& {Weiss}}]{Miller}
{Miller Bertolami}, M.~M., {Althaus}, L.~G., {Unglaub}, K., \& {Weiss}, A.
  2008, \aap, 491, 253 \csname Millerlink\endcsname~\csname
  Millernote\endcsname

\bibitem[{{Moran} {et~al.}(1999){Moran}, {Maxted}, {Marsh}, {Saffer}, \&
  {Livio}}]{Moran}
{Moran}, C., {Maxted}, P., {Marsh}, T.~R., {Saffer}, R.~A., \& {Livio}, M.
  1999, \mnras, 304, 535 \csname Moranlink\endcsname~\csname
  Morannote\endcsname

\bibitem[{{Napiwotzki} {et~al.}(2004){Napiwotzki}, {Karl}, {Lisker}, {Heber},
  {Christlieb}, {Reimers}, {Nelemans}, \& {Homeier}}]{Napiwotzki}
{Napiwotzki}, R., {Karl}, C.~A., {Lisker}, T., {et~al.} 2004, \apss, 291, 321
  \csname Napiwotzkilink\endcsname~\csname Napiwotzkinote\endcsname

\bibitem[{{Naslim} {et~al.}(2012){Naslim}, {Geier}, {Jeffery}, {Behara},
  {Woolf}, \& {Classen}}]{Naslim}
{Naslim}, N., {Geier}, S., {Jeffery}, C.~S., {et~al.} 2012, \mnras, 423, 3031
  \csname Naslimlink\endcsname~\csname Naslimnote\endcsname

\bibitem[{{N{\'e}meth} {et~al.}(2012){N{\'e}meth}, {Kawka}, \&
  {Vennes}}]{Peter}
{N{\'e}meth}, P., {Kawka}, A., \& {Vennes}, S. 2012, \mnras, 427, 2180 \csname
  Peterlink\endcsname~\csname Peternote\endcsname

\bibitem[{{Orosz} \& {Wade}(1999)}]{Orosz}
{Orosz}, J.~A. \& {Wade}, R.~A. 1999, \mnras, 310, 773 \csname
  Oroszlink\endcsname~\csname Orosznote\endcsname

\bibitem[{{Paxton} {et~al.}(2011){Paxton}, {Bildsten}, {Dotter}, {Herwig},
  {Lesaffre}, \& {Timmes}}]{Paxton2011}
{Paxton}, B., {Bildsten}, L., {Dotter}, A., {et~al.} 2011, \apjs, 192, 3
  \csname Paxton2011link\endcsname~\csname Paxton2011note\endcsname

\bibitem[{{Paxton} {et~al.}(2013){Paxton}, {Cantiello}, {Arras}, {Bildsten},
  {Brown}, {Dotter}, {Mankovich}, {Montgomery}, {Stello}, {Timmes}, \&
  {Townsend}}]{Paxton2013}
{Paxton}, B., {Cantiello}, M., {Arras}, P., {et~al.} 2013, \apjs, 208, 4
  \csname Paxton2013link\endcsname~\csname Paxton2013note\endcsname

\bibitem[{{Paxton} {et~al.}(2015){Paxton}, {Marchant}, {Schwab}, {Bauer},
  {Bildsten}, {Cantiello}, {Dessart}, {Farmer}, {Hu}, {Langer}, {Townsend},
  {Townsley}, \& {Timmes}}]{Paxton2015}
{Paxton}, B., {Marchant}, P., {Schwab}, J., {et~al.} 2015, \apjs, 220, 15
  \csname Paxton2015link\endcsname~\csname Paxton2015note\endcsname

\bibitem[{{Saffer} {et~al.}(1998){Saffer}, {Livio}, \&
  {Yungelson}}]{Saffer1998}
{Saffer}, R.~A., {Livio}, M., \& {Yungelson}, L.~R. 1998, \apj, 502, 394
  \csname Saffer1998link\endcsname~\csname Saffer1998note\endcsname

\bibitem[{{Schindler} {et~al.}(2015){Schindler}, {Green}, \&
  {Arnett}}]{Schindler}
{Schindler}, J.-T., {Green}, E.~M., \& {Arnett}, W.~D. 2015, \apj, 806, 178
  \csname Schindlerlink\endcsname~\csname Schindlernote\endcsname

\bibitem[{{Sweigart} {et~al.}(2004){Sweigart}, {Lanz}, {Brown}, {Hubeny}, \&
  {Landsman}}]{Sweigart}
{Sweigart}, A.~V., {Lanz}, T., {Brown}, T.~M., {Hubeny}, I., \& {Landsman},
  W.~B. 2004, \apss, 291, 367 \csname Sweigartlink\endcsname~\csname
  Sweigartnote\endcsname

\bibitem[{{Vos} {et~al.}(2015){Vos}, {{\O}stensen}, {Marchant}, \& {Van
  Winckel}}]{Vos2015}
{Vos}, J., {{\O}stensen}, R.~H., {Marchant}, P., \& {Van Winckel}, H. 2015,
  \aap, 579, A49 \csname Vos2015link\endcsname~\csname Vos2015note\endcsname

\bibitem[{{Wang} \& {Han}(2009)}]{Wang}
{Wang}, B. \& {Han}, Z. 2009, \aap, 508, L27 \csname Wanglink\endcsname~\csname
  Wangnote\endcsname

\end{thebibliography}

\end{document}